\documentclass[prx, 10pt, twocolumn, floatfix, groupedaddress] {revtex4-1}

\usepackage{times, mathrsfs, amsmath, amsfonts, graphics, graphicx, cancel, color, theorem, bbm, mathtools, amssymb}

\usepackage[nice]{nicefrac}
\usepackage[english]{babel}
\usepackage[margin=.75in]{geometry}
\usepackage[T1]{fontenc}
\usepackage{tabularx,ragged2e,booktabs}
\usepackage{capt-of}
\newcolumntype{C}[1]{>{\Centering}m{#1}}

\usepackage{xfrac,relsize,float}

\usepackage{appendix}

\usepackage[unicode=true, breaklinks=false, pdfborder={0 0 1}, backref=false, colorlinks=true, linkcolor=blue, citecolor=blue]{hyperref}

\def\BraVert{\egroup\,\mid\,\bgroup}

\def\KetBra#1#2{\left|#1\vphantom{#2}\right\rangle\!\!\left\langle\vphantom{#1}#2\right|}

\definecolor{Blue}{rgb}{0,0,1}
\definecolor{Red}{rgb}{1,0,0}
\definecolor{Green}{rgb}{0,1,0}
\definecolor{darkgreen}{rgb}{0,.7,0}
\definecolor{Purp}{rgb}{.2,0,.2}
\definecolor{white}{rgb}{1,1,1}

\newcommand{\tr}{{\rm tr}}

\begin{document}
\title{Non-Markovian memory in IBMQX4}
\author{Joshua Morris}
\email{josh.morris@monash.edu}
\author{Felix A. Pollock}
\email{felix.pollock@monash.edu}
\author{Kavan Modi}
\email{kavan.modi@monash.edu}
\affiliation{School of Physics \& Astronomy, Monash University, Clayton, Victoria 3800, Australia}
\date{\today}

\begin{abstract}
We measure and quantify non-Markovian effects in IBM's Quantum Experience. Specifically, we analyze the temporal correlations in a sequence of gates by characterizing the performance of a gate conditioned on the gate that preceded it. With this method, we estimate (i) the size of fluctuations in the performance of a gate, i.e., errors due to non-Markovianity; (ii) the length of the memory; and (iii) the total size of the memory. Our results strongly indicate the presence of non-trivial non-Markovian effects in almost all gates in the universal set. However, based on our findings, we discuss the potential for cleaner computation by adequately accounting the non-Markovian nature of the machine.
\end{abstract}

\maketitle

\section{Introduction}

Quantum computers promise significant speedups over their classical counterparts by manipulating and exploiting effects only seen in the quantum realm~\cite{qsimulation, qc_experiment, fsimulation}. The price of using phenomena that only become significant at the smallest scales is incredible sensitivity to noise and inherent fragility~\cite{expo_sensitivity, rb_exp}. Quantifying this fragility, in general, necessitates sophisticated methods for understanding and predicting the behaviour of the fundamental constituents of computing, i.e., quantum gates~\cite{rb_actual, GST_PRA, rb_wallman}.

Imperfections in quantum gates arise, in part, from interactions with an uncontrollable environment. Any realistic quantum computer is an open system, but the nature of the interactions with the environment are not always clear. In the broadest sense, the resulting dynamics can be partitioned into one of two categories, Markovian and non-Markovian~\cite{nonMarkovPRL, arxiv:1512.07106}, depending on whether memory effects play a role. In the former case, information leaks out of the system over time, eventually reaching an equilibrium point, beyond which further computation becomes impossible or meaningless. In the non-Markovian regime, the information transfer between the system and its environment becomes bidirectional. At some later time information lost to the environment may return to the system, resulting in behaviour that depends on the system's previous state. While, in some cases, these temporal correlations could be beneficial~\cite{harness_nonmark}, without knowledge of their specific structure, they manifest as unwanted noise. This noise furthermore violates many common assumptions made in characterizing and controlling it~\cite{m_NM_correction}.

In a circuit-based quantum computer, where sequential gates are applied to realize the computational steps of a specific algorithm, if errors were to non-trivially depend on previous choices of gates, meaningful computation would become impossible. While quantum error correction could certainly be employed to minimize these effects, most error correction techniques almost always assume that the errors are Markovian and thus correct sub-optimally for correlated errors; see~\cite{Aharonhov_lr, preskill_scalableN, noise_kalai} for exceptions. Moreover, techniques for gauging the performance of a quantum computer, such as randomized benchmarking~\cite{rb_exp} and gate set tomography~\cite{GST_PRA}, fail to be reliable in the presence of non-Markovian errors~\cite{RB_correlations, rb_wallman}.

Recently, several quantum computers have been made available to researchers to run experiments remotely, with IBM's Q Experience a prominent example. Researchers have used IBM's quantum computers to prepare highly entangled states~\cite{arXiv:1801.03782}, discriminate between unitary operations~\cite{arXiv:1807.00429}, implement quantum stochastic differential equations~\cite{qdiff_qc}, test fault-tolerant protocols~\cite{IBM_fault, arXiv:1807.02883} and demonstrate dynamical decoupling~\cite{IBMStateFid}. We add to this growing list here.

In this article, we show evidence for the existence of temporal correlations between sequential gates implemented on the IBMQX4, a five superconducting transmon qubit quantum computer (dubbed Tenerife)~\cite{gambetta}. To demonstrate non-Markovian effects in the IBMQX4, we develop techniques for quantifying the conditional dependence of noise in quantum gates on the history of past operations and identifying the approximate time scales of the corresponding correlations, all without specific information about the underlying system-environment interactions. Specifically, we look at the behaviour of a quantum gate conditioned on the gate that precedes it, finding noise that depends on past choices and hence a strong indication of non-Markovianity. Finally, we estimate the size of the memory by measuring the correlations in a sequence of controlled-not gates. Our findings indicate that the IBMQX4 is non-trivially coupled to its environment, and suffers from non-Markovian effects that cannot be ignored. On the other hand, our results could, in turn, be used to inform the design of better pulse sequences conditioned on previous pulses to either sidestep or mitigate the issue~\cite{qcomp_soft}, yielding cleaner and more faithful computation.

We begin with an overview of our theoretical ideas, followed by presentation of data found by running remote experiments on the IBMQX4.

\begin{figure}[ht]
\includegraphics[width=0.45\textwidth]{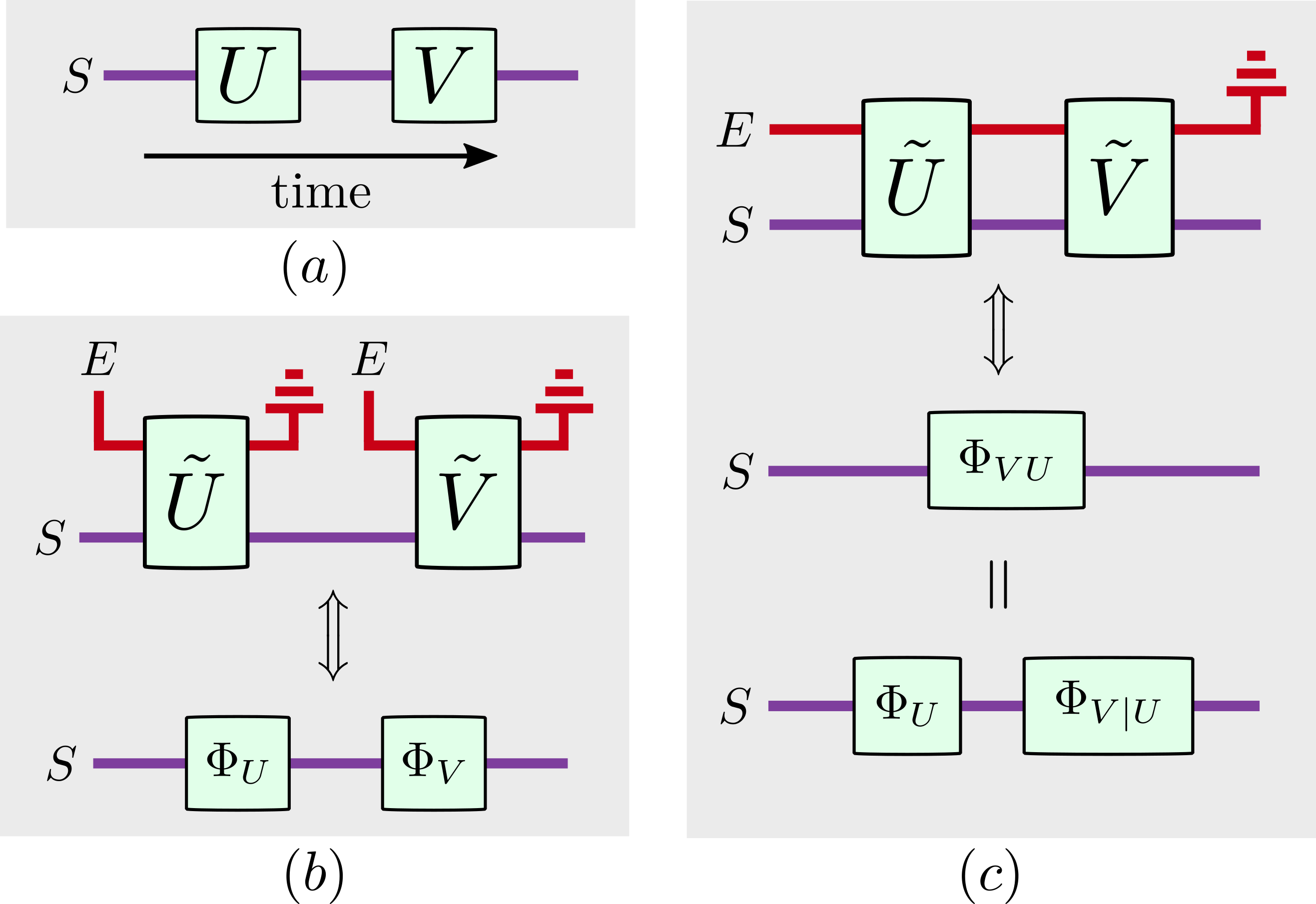}
\caption{(a) Two unitary operators act on closed system $S$, evolving it in time. (b) If the system evolves alongside an inaccessible bath or environment $E$, but does not affect it sufficiently to impact on future interactions, then it is Markovian (or can be approximated as such) and the dynamics on $S$ alone will be described by a CPTP map. (c) Otherwise, the system is non-Markovian,  and the environment can be thought of as an uncontrollable quantum memory attached to the system. While the collective dynamics on $SE$ will be a sequence of unitaries, the local dynamics of $S$ cannot be straightforwardly decomposed into two independent stages. Note that each of the boxes in these panels should be thought of as a quantum map (sometimes labelled by a unitary operator) acting on a density operator.\label{fig:ideal}}
\end{figure}

\section{Testing for non-Markovianity}

Consider the three scenarios outlined in Fig.~\ref{fig:ideal} for the sequential application of two quantum gates. The first panel represents the desired evolution; that of a closed quantum system undergoing unitary transformations. In reality, the implementation of a quantum gate $U$ is unlikely to be perfect; instead, a noisy operation, described by a completely-positive, trace-preserving (CPTP) map or quantum channel $\Phi_U$~\cite{MilzReview}, will be applied to the system. We may think of $\Phi_U$ as stemming from a unitary transformation $\tilde{U}$ applied on the system-environment. When applying an operator $U$ followed by another operator $V$, it is often assumed that this leads to application of $\Phi_U$ followed by $\Phi_V$. That is, the two noisy maps are independent, as depicted in Fig.~\ref{fig:ideal}b.

However, this assumption of independence between applications of quantum gates requires the information carried by the environment to dissipate from one application to the next. This Markovian behaviour is in contradistinction to the circuit in Fig.~\ref{fig:ideal}c, where the environment evolves \emph{without} complete information loss, leading to a noisy process 
\begin{gather}
    \Phi_{VU} \ne \Phi_{V} \circ \Phi_{U}.
\end{gather}
The resulting non-Markovian dynamics leads to correlations between gate errors, which should be accounted for in a rigorous analysis of fault tolerance.

To quantify these correlations, we exploit the fact that in the Markovian case, with no temporal correlations between gates, $\Phi_{VU}$ would be equivalent to the composition of individual maps $\Phi_{V} \circ \Phi_U$. Let us define the conditional map
\begin{gather}
    \Phi_{V|U} := \Phi_{VU} \circ \Phi_U^{-1}, \quad \mbox{where} \quad \Phi_U \circ \Phi_U^{-1}= \mathcal{I},
\end{gather}
with $\mathcal{I}$ being the identity map ($\mathcal{I}[\rho] = \rho\; \forall \rho$). If $\Phi_{V|U}$ is completely positive, then the dynamics is what is called CP-divisible~\cite{simon-div}, and this is will be our first test for non-Markovian memory. Even when the conditional map is CP, this alone does not guarantee a Markov process, since the conditional map can still depend on its conditioning argument, and a broader analysis technique is required.

For our second check we ask how different is $\Phi_{V|U}$ from $\Phi_V$. If the two are not the same, this implies non-Markovianity. However, for a skeptic this failure could imply a very mild type of non-Markovianity. That is, suppose the performance of the second gate is always worse than the first gate, i.e., 
\begin{gather}
    \Phi_{VU} = \Phi_V \circ \Phi_2 \circ \Phi_U
\end{gather}
for some fixed decohering dynamics $\Phi_2$ associated with application of any second gate. This is a simple non-Markovian process, where the memory is merely a clock keeping track of the number of gates that have been applied. To overcome this we further implement a more sophisticated check; a less trivial form of non-Markovianity would manifest as explicit dependence on the first gate, i.e., 
\begin{gather}
    \Phi_{V|U_1} \ne \Phi_{V|U_2} \quad \mbox{for all} \quad U_1,\ U_2.
\end{gather}
We now construct all three of these tests for all gates in $\mathcal{G}$, a universal set of gates.

\begin{figure}[t]
  \centering
  \includegraphics[width=0.3\textwidth]{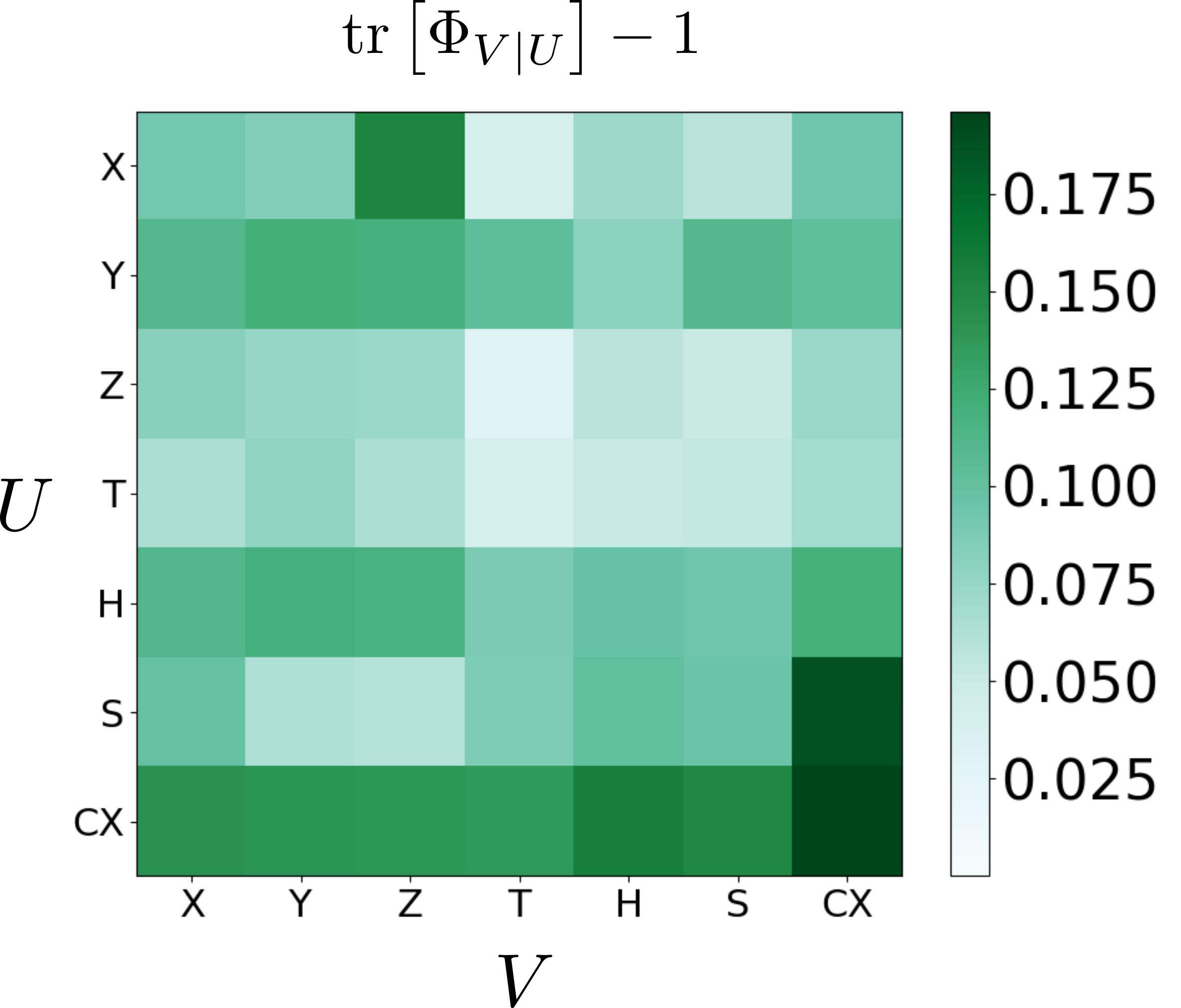}
    \caption{As a measure of the non-positivity of the reconstructed inverse maps $\Phi_{V|U}$, we present the difference from unity of the summed absolute values of the maps' Choi states (operator representations of the maps~\cite{MilzReview}): $\tr|\Phi_{V|U}| - 1$ for each of the gate combinations $U,V \in \mathcal{G}$. For completely positive quantum maps this should be vanishing. The deviation of this measure from zero then gives the degree to which the reconstructed map is not CP, indicating non-Markovian dynamics.} 
  \label{fig:NCP}
\end{figure}

\section{Conditional performance}

To clearly demonstrate the presence of temporal correlations, we implement single and sequential pairs of gates 
\begin{gather}\label{eq:G}
U,V \in \mathcal{G} = \{H, S, T, X, Y, Z, C_X\}
\end{gather}
on the IBMQX4. Here $H$ is the Hadamard gate, $S$ is the phase gate, $T$ is the $\nicefrac{\pi}{8}$ gate, $X,Y,Z$ are the Pauli gates, and $C_X$ is the \textsc{Cnot} gate. We reconstruct, via quantum process tomography (QPT), the maps $\{\Phi_U, \Phi_{VU}\}$ corresponding to each one- and two-gate sequence, i.e., $\{\Phi_U, \Phi_{VU}\}$. The details of QPT are given in the Appendix~\ref{appendix:ptom}, along with an analysis of the associated statistical and systematic errors in Appendix~\ref{app:B}. 

With the relevant maps reconstructed, we perform three tests described above for the presence of temporal correlations. The first is a simple check of complete positivity of each conditional map  $\Phi_{V|U}$, computed from the corresponding  $\Phi_{VU}$ and  $\Phi_{U}$. All of the gate combinations tested fail to satisfy this condition, as seen in Fig.~\ref{fig:NCP}; however, as mentioned above, even if they were completely positive, this would not rule out non-Markovianity. We therefore move on to the more sophisticated second and third tests, which involve direct comparison of each $\Phi_{V}$ with various $\Phi_{V|U}$, and of pairs $\Phi_{V|U_1}$ and $\Phi_{V|U_2}$, respectively.

There are many ways for comparing two quantum maps, with the natural metric being the diamond distance~\cite{alexei_metric,benenti_diamond}
\begin{gather}\label{eq:diamond}
\|\Phi_A - \Phi_B \|_\diamond := \max_{\xi} D(\Phi_A\otimes \mathcal{I}_d[\xi], \Phi_B\otimes \mathcal{I}_d[\xi]),
\end{gather}
where 
\begin{gather}\label{eq:dist}
D(\rho,\sigma) = \tfrac{1}{2}{\rm tr}|\rho-\sigma|
\end{gather}
is the trace distance, $\mathcal{I}_{d}$ is the identity map on an ancillary $d$ dimensional space, and $\rho$ is a density operator on the joint system-ancilla space. This distance can be interpreted as the maximum ability to statistically distinguish between two maps in a single shot with an entangled input.
However, the diamond distance in our context accounts for the worst case distinguishability; a given pair of maps are unlikely to differ this much when applied to a typical input. In addition, we would like a measure that characterises the average distinguishability between two maps. That is, we want to quantify the effects of non-Markovianity a typical user of IBMQX4 might encounter. To achieve this we employ the averaged trace distance

\begin{gather}\label{eq:cond_dist}
E(\Phi_A,\Phi_B) := \frac{1}{M}\sum_{l=1}^M {D(\Phi_A[\rho_l], \Phi_B[\rho_l ])},
\end{gather}
where $\{\rho_l\}_{l=1}^M $ forms a set of $M$ random input states drawn from the Haar distribution \cite{state_survey}.

\begin{figure}[t]
\centering
\includegraphics[width=0.48 \textwidth]{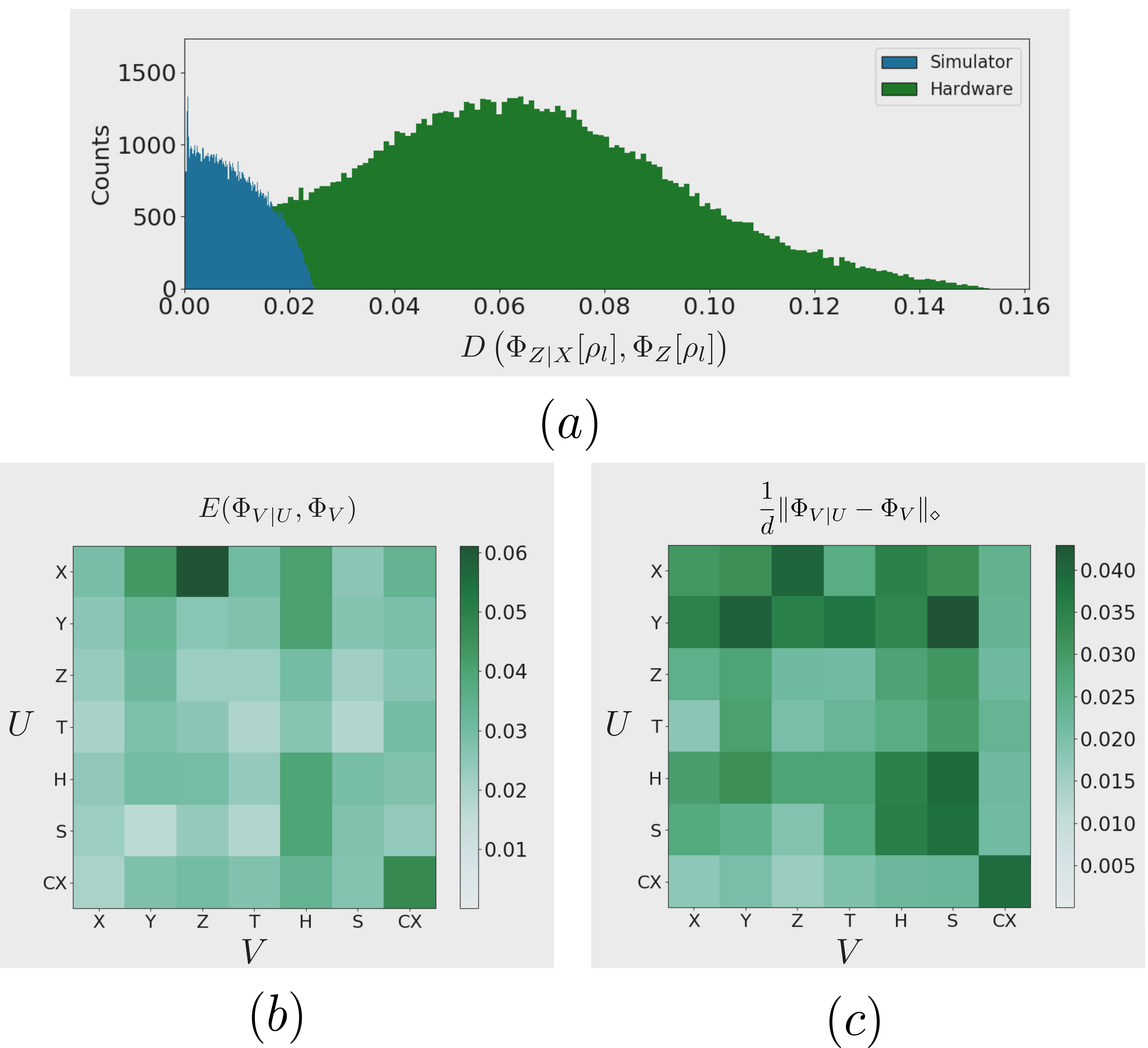}
\caption[Correlated operations]{(a) The two maps $\Phi_{Z|X}$ and $\Phi_Z$ are found via process tomography on the IBMQX4. Both maps are applied to one hundred thousand randomly generated input states, and their output states are compared against one another (Green) using Eq.~\eqref{eq:dist}. The same quantities computed using a simulator provided by IBM are also presented (Blue). (b) The average trace distance between outputs of $\Phi_{V|U}$ and $\Phi_V$ for each combination of $U,V\in \mathcal{G}$, as given by Eq.~\eqref{eq:cond_dist}. (c) The diamond distance, given in Eq.~\eqref{eq:diamond}, between the same pairs of maps, but scaled by $\tfrac{1}{d}$, with $d=4$ when $U$ or $V$ is $C_X$ and $d=2$ otherwise. This may be thought of (up to this scaling) as calculating the supremum of the distribution in (a), whereas (b) gives its mean. The uncertainty in these values is approximately $4.5\times 10^{-3}$. Note that for a constant operator $V$, both metrics fluctuate as $U$ is varied, indicating a dependency on the previous gate.}
 \label{fig:errors}
\end{figure}

In Fig.~\ref{fig:errors}a, we show the distribution of trace distances, prior to averaging, for the gate sequence $U=X, V=Z$. The figure shows this distribution for both experimentally determined maps, as well as maps constructed using the IBMQX4 simulator. The latter only accounts for statistical errors, while the former is equipped to account for memory effects. The figure shows that IBMQX4 suffers from systematic errors that go beyond statistical fluctuations. The average of this distribution indicates size of error, on the average, due to the non-Markovianity.

In Figs.~\ref{fig:errors}b and c, we show the distances, according to Eq.~\eqref{fig:errors} and Eq.~\eqref{eq:diamond} respectively, between conditional and unconditional maps corresponding to gates in the set $\mathcal{G}$ applied on IBMQX4. To maintain consistency, each single qubit gate always acts on qubit~1 (with numbering from~0-4) and each controlled-not ($C_X$) acts on qubit~0 with the control on qubit~1. This analysis clearly shows the conditional realization of the gates to be different from that of unconditional ones. This might be expected, as the coherence of the qubit will diminish more after two operations in comparison to just one. In this case, the magnitude of each column of the matrix entries in Figs.~\ref{fig:errors}b and c would be the same. But they are not: for instance, performing a $Z$ gate after an $X$ gate is very different than after another $Z$. This significant variation as the initial operation $U$ is varied, for a particular $V$, is precisely the dependence on the past discussed above. Beyond simply witnessing the existence of non-Markovian errors in the IBMQX4, Figs.~\ref{fig:errors}b and c tell us which \textit{specific} gates lead to the strongest (detectable) interaction with the environment.

\begin{figure}[t]
 \centering
 \includegraphics[width=0.45\textwidth]{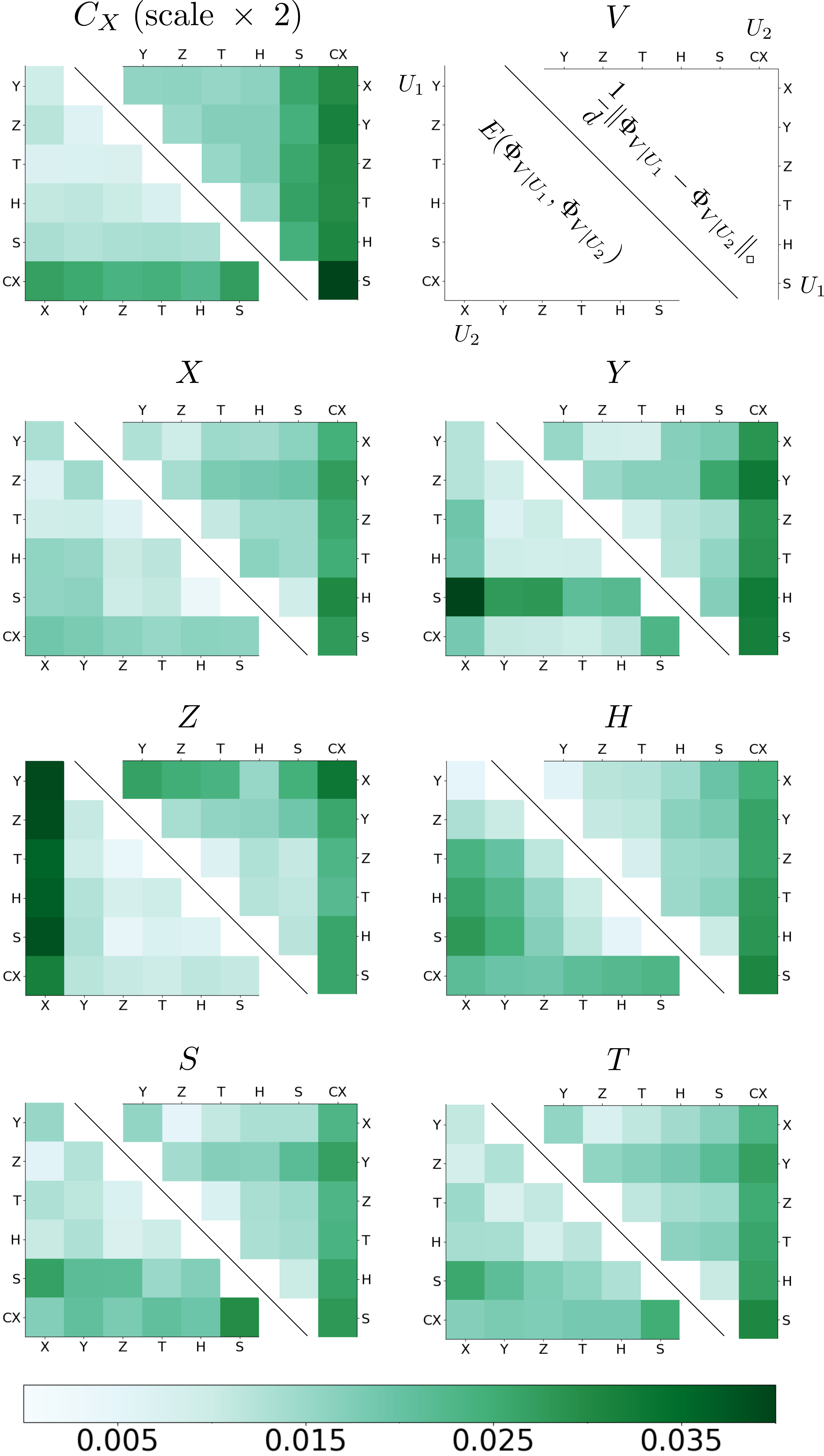}
\caption{The distinguishability between $\Phi_{V|U_1}$ and $\Phi_{V|U_2}$ for all $V, U_1, U_2 \in \mathcal{G}$ using the diamond distance (upper triangles) and average trace distance (lower triangles). The former is scaled by a factor of $\tfrac{1}{d}$ ($d=4$ for $C_X$, $d=2$ otherwise), and both measures are scaled by a factor of two for $V = C_X$.} 
  \label{fig:conditionals}
\end{figure}

Though this result demonstrates some kind of correlation in sequential processes, it allows for the comparison of the effect of past gate choices only indirectly; two different choices of $U$ might lead to $\Phi_{V|U}$s that are similarly distinguishable from $\Phi_V$, but which are also significantly distinct from each other. For a more direct comparison, we compute the distingushability between maps corresponding to a fixed gate conditioned on different preceding gates: $E(\Phi_{V|U_1}, \Phi_{V|U_2})$. This is shown in Fig.~\ref{fig:conditionals} for $V, \,U_1, \ U_2 \in \mathcal{G}$. It can be seen that the difference between conditional maps is far from uniform. In other words, a gate's deviation from its ideal implementation is significantly perturbed by past actions, and not with others, when implemented on the IBMQX4. For instance, $Y$ is strongly affected by $S$, but less so by other gates. Similarly, $Z$ seems to be most affected only by $X$, while $X$ is sensitive to any gate that precedes it. That is, the behaviour of $X$ is drastically different conditionally on what came before it. These structures strengthen the confidence in our analysis.

As important as their presence is, the lifetime of these correlations may be relatively short. Even strong correlations between sequential actions can be accounted for if their temporal range is small; one need only wait them out before application of the next gate. It is with this in mind that we proceed to an investigation of the lifetime of this newly detected quantum memory.

\section{Memory length}

Having showed evidence for temporal correlations between pairs of sequential operations on the IBMQX4 platform, we now ask how long-lived they are. In addition to waiting between gate applications, if the length of the memory is not too long, then a hidden Markov model could be reconstructed and conditional pulse sequences applied to correct for errors~\cite{phil_QMO, phil_finite}. If, however, these correlations extend far into the system's future, then correcting or modelling them may be challenging. Additionally, having a better grasp of the structure of memory would also enable more informed decoupling techniques.

To estimate the length of the memory in IBMQX4, we apply a sequence of $n$ $C_X$ gates and construct the corresponding CPTP maps $\{\Phi^{(n)}_{C_X}\}$ using process tomography for each $n \in [1,15]$. In the Markovian case, where the errors in each implementation are independent, we would expect $E(\Phi^{(n)}_{C_X}, \Phi^{(m)}_{C_X} \circ \Phi^{(n-m)}_{C_X})$ to vanish for integers $m<n$. Conversely, as before, a non-vanishing distinguishability, by either measure we consider, is a measure for temporal correlations~\cite{simon-div}. Furthermore, the behaviour as a function of $m$, for fixed $n$, indicates whether and how the memory decays. Rather than the average distinguishability of two maps that we consider above, the diamond distance considers the absolute worst case scenario by maximising the trace distance between map outputs over all inputs, including those that include entangled ancillas. This maximisation problem can be posed as a semi-definite program~\cite{dnorm_watrous}, described and solved using an appropriate software package~\cite{cvxpy,cvxpy_rewriting}.

\begin{figure}
\centering
 \includegraphics[width=0.45\textwidth]{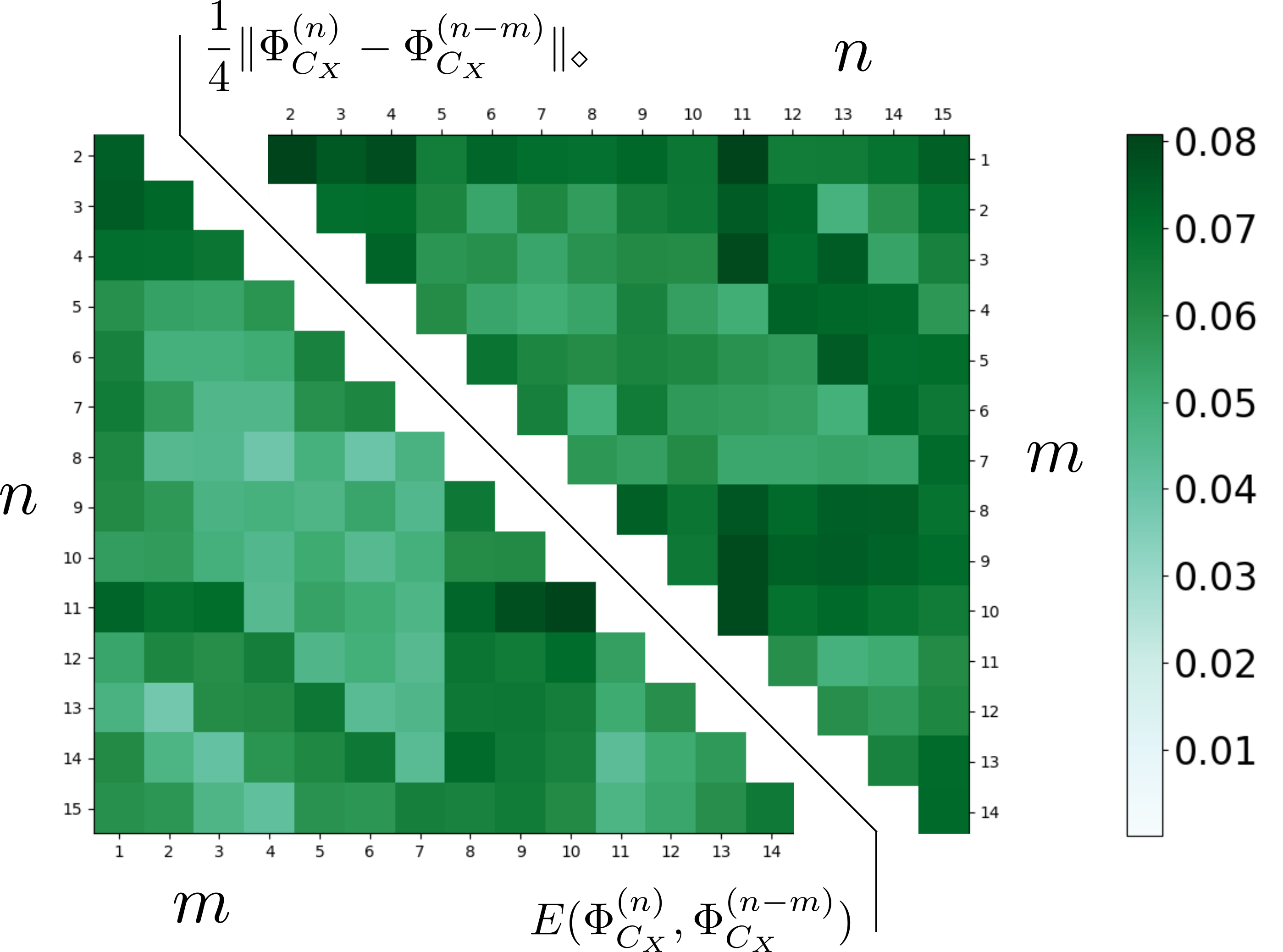}
 \caption[Memory length]{The distingushability of two maps; one constructed from the application of $n$ sequential $C_X$ gates $\Phi_{C_X}^{(n)}$, and the other a concatenation of maps corresponding to $m$ and $n-m$ sequential $C_X$ gates $\Phi_{C_X}^{(m)} \circ \Phi_{C_X}^{(n-m)}$. In the upper figure the distinguishability is calculated using the diamond distance in Eq.~\ref{eq:diamond}, while in the lower one it is computed as in Eq.~\eqref{eq:cond_dist}.}
 \label{fig:memory_decay}
\end{figure}

Unsurprisingly, given our previous results, we see significant fluctuations in the distinguishability as $m,n$ are varied, shown in Fig.~\ref{fig:memory_decay}. Though we are now considering the absolute error of $\Phi^{n}$ and the causally broken process $ \Phi^{(m)} \circ \Phi^{(n-m)} $, as opposed to the relative error of two processes as in Fig.~\ref{fig:errors}b, we are still comparing the two maps as they actually are in the IBMQX4, rather than the ideal case. This is to say that Fig.~\ref{fig:memory_decay} is not intended as a commentary on the fidelity of the $C_X$ gate in the IBMQX4, but rather its operational dependence on the environmental state after repeated applications of the controlled not gate.

From Fig.~\ref{fig:memory_decay}, it is clear that, beyond the lack of divisibility, there is some structure present in the distingushability as the temporal length of the two maps is increased. Considering the upper triangular matrix of the figure, where we compute the diamond distance between the two channels; we see a high channel distinguishability, corresponding to strong short term correlations, before a decrease with increasing duration. This decrease is presumably due to the decoherence of both maps as they converge on a noisy equilibrating channel. At $m=8$, however, the divergence once again increases, indicating that some kind of long range temporal correlation has been lost due to the imposed causal break. Given that the $C_X$, as a coupling operation, acts on a larger physical space than local operations, it is not surprising that we see evidence for temporal correlations. Curiously, after $m=10$, the distinguishability once again drops quite sharply, where we might instead expect a gradual decrease. While this behaviour for $m=8,9,10$ could indicate a different calibration of the device during their reconstruction, all tomography experiments for related maps in Fig.~\ref{fig:memory_decay} were performed at similar times on the IBMQX4.

Our results suggest that the non-Markovian memory lasts for several gates. Though it is not clear what physical mechanism corresponds to these long terms correlations, we find it encouraging that we can identify them with no knowledge of the corresponding gate implementations on the IBMQX4. Combining our results with a physical model will open up the potential for a hidden Markov model for IBMQX4. We now move to discuss broader implications of our results.

\section{Full quantification of non-Markovianity}

\begin{figure}[t]
\centering 
 \includegraphics[width=0.35\textwidth]{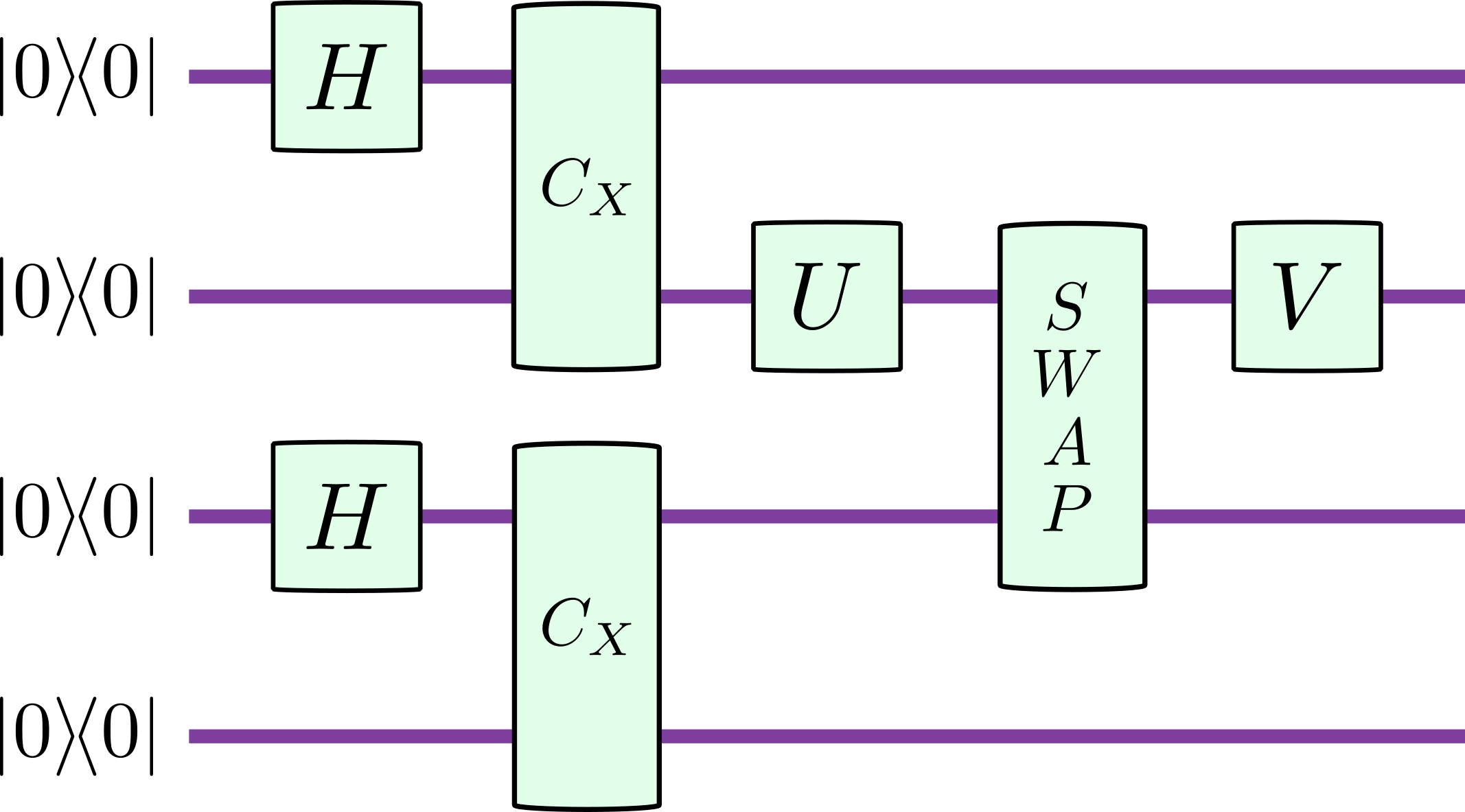}
 \caption[Correlated operations]{The quantum circuit that would create the Choi state of the process tensor corresponding to sequential application of gates $U$ and $V$, assuming all other gates are applied cleanly. For Markovian noise in $U$ and $V$, this would just be the product of Choi states of their corresponding maps $\Phi_U$ and $\Phi_V$.}
 \label{fig:cji}
\end{figure}

Though we have presented measures for correlations between errors in gate applications at different points in time, this falls short of a full characterization of the corresponding non-Markovian process. Ideally, we would reconstruct the \textit{process tensor}~\cite{nonMarkov, arxiv:1512.07106}, which is a generalization of the CPTP map to multi-step processes. The process tensor is a complete descriptor of a stochastic quantum process\cite{kolmogorov, arXiv:1702.01845}, including  temporally correlated noise, which could be of purely quantum nature~\cite{arXiv:1710.01776, arXiv:1811.03722}. However, reconstructing the process tensor requires sequential measurements, which are currently not possible in the IBMQX4 system. A potential way around this is to construct a restricted process tensor~\cite{PhysRevA.98.012108}, but it too can be cumbersome. Another alternative is to map the process to the large entangled quantum state  realized by the circuit depicted in Fig.~\ref{fig:cji}, which physically implements a generalized Choi-Jamio\l{}kowski isomorphism. This could then be determined through state tomography. Though more resource intensive in terms of qubits required, this procedure does not require sequential measurements.

Unfortunately, even this cannot be achieved, since implementing it requires the clean implementation of $H$, $C_X$ and swap (a sequence of three $C_X$) gates, which we have seen are themselves noisy and temporally correlated. These correlations and the imperfect nature of the gates means we construct not the Choi state of the process tensor but rather some proxy for it. Determining this proxy experimentally however can still yield meaningful results. We have run the circuit in Fig.~\ref{fig:cji}, for the choices of $U=S$ and $V=T$, on the older IBMQX2 and performed quantum state tomography on the relevant four qubits. We then compared this with a Markovian simulator of the same circuit. This is realized by multiplying the reconstructed map for each element of the circuit numerically. The relative entropy distance between the two states represents an estimate of the amount of non-Markovianity~\cite{nonMarkovPRL, arxiv:1512.07106}, which for IBMQX2 we find to be $0.68$ (with $2$ being the maximum achievable~\footnote{Relative entropy for four qubits can be as large as 4, but due to causality conditions this is not achievable for the process tensor.}). Though this value cannot be taken as a measure of the correlation between the maps corresponding to gates $U$ and $V$ in Fig.~\ref{fig:cji}, it can serve as an indicator and measure of non-Markovianity, albeit with little insight into the source. It is worth noting that IBMQX2 appears to be much noisier than IBMQX4~\cite{arXiv:1805.07185}. Nevertheless, our findings suggest that a great deal of quantum information may be trapped in the non-Markovian correlations in both devices. Being able to recover this information would make the quantum computation cleaner. 

Our findings are supported by previous work, such as Ref.~\cite{IBMStateFid}. There the authors focus on constructing quantum dynamics that are resistant to noise through dynamical decoupling. They experimentally implement a dynamical decoupling protocol on both the IBMQX4 and the Rigetti Acorn (a 19 superconducting qubit computer). They demonstrate enhancement in the fidelity of operations in the IBMQX4 due to their decoupling protocol, which is a purely a non-Markovian phenomenon~\cite{sakuldee_2018, li_concepts_2018}.

\section{Discussions}

Although, we cannot fully quantify non-Markovianity in IBMQX4, we have presented two techniques for identifying and measuring the properties of a non-Markovian system. These techniques form a useful tool-set in identifying difficult to detect correlations in specific, sequential control operations for the IBMQX4 and for general quantum systems where process tomography (or some analogue of it) is possible. Our methods require no prior knowledge of the system Hamiltonian to infer the temporal correlations and as such we need not have an in-depth understanding of the experimental details of the IBMQX4 to comment on its physical behaviour. On the other hand, combining our methods with a physical model would allow for better estimates for the physical parameters, e.g., the coupling strengths, using machine learning tools~\cite{arXiv:1704.00800, arXiv:1901.05158}. Moreover, through an iterative procedure our could be used to make a better physical model for IBMQX4, and consequently adjust the pulse sequence to yield cleaner gates.

The hurdle that we face in reconstructing the process tensor is also a main drawback in reconstructing the conditional dynamics. The root cause being the gate infidelity in what is essentially the preparation and measurement phases of tomography, i.e., state preparation and measurement (SPAM) errors. In performing process tomography we require the preparation of specific input states to the gate being reconstructed and specific measurements performed on the output state. A convenient assumption for most tomography~\cite{GST_rev} is that these preparations and measurements are performed flawlessly, with the only error being statistical in nature (stemming from finite sampling of the outcome distribution). If this is not the case, however, then the determination of maps corresponding to different gates, and their comparison, can become unreliable. The error rates, based on randomized benchmarking, for IBMQX4~\cite{IBM_website} suggest that SPAM errors by themselves are small. Moreover, the SPAM errors alone are not sufficient to explain the high level of structure and the asymmetry in Figs.~\ref{fig:errors}, ~\ref{fig:conditionals}, and~\ref{fig:memory_decay}. For the complete treatment of these errors see Appendix~\ref{app:B}.

Putting aside the IBMQX4, the ability to detect gate specific correlations, as we have demonstrated above, is a useful diagnostic tool when designing quantum devices. Though a single quantum gate may be well-implemented in isolation, the act of applying it may cause unwanted and traditionally difficult to detect perturbations in the next gate. Our techniques are designed to address this exact situation. In addition to this, though we have considered only temporal dependence mediated by some external environment here, this technique could easily be extended to a search for spatio-temporal correlations, e.g. find the correlations between the third gate applied to qubit one and the fifth gate applied to qubit six. Besides measuring the performance of the device, in a way that is not biased by its particular architecture, this confers a number of advantages. In particular, we note the possibility of the results shown here being used to identify where efforts to improve the implementation of gates might be best spent. At the very least, our results highlight where non-Markovian correlations play a role, allowing for improved constraints on meaningful quantum circuit compilation, something that is, and will continue to be, a problem for large scale quantum computation~\cite{qcompiler, qcomp_soft}.

\appendix

\section*{Appendices}

\section{Process tomography in the IBMQX4}
\label{appendix:ptom}

Due to our need to determine and compare the transformations that are actually being performed in the IBMQX4, it was required that we perform process tomography on a number of one and two step gate sequences. In the IBM system, only Pauli $Z$ measurements are possible, thus a rotation operation needs to be performed for measurement in a complete tomographic basis (the minimal Pauli basis set in this instance). Since the computer is an open system, Markovian or otherwise, we are limited to determining the reduced dynamics of the full unitary evolution of the system-environment. This is described by the quantum map $\Phi[\rho_S] = \tr_E(\tilde{U}\rho_{SE}\tilde{U}^\dagger)$ that acts on $\rho_S$ and may be expressed in tomographic form, as presented in~\cite{MilzReview}:
\begin{gather}\label{eq:tomog}
\Phi[\rho] =  \sum_i \rho_i^\prime \tr \left( D_i^\dagger \rho \right),
\end{gather}
where $\rho_i^\prime = \Phi[\rho_i]$ is the output state of the map and $\mathcal{S} = \{\rho_i \}_{i=1}^{d^2}$ spans the space of bounded linear operators $\mathcal{B}(\mathcal{H}_d)$ acting on the system Hilbert space (with dimension $d$). $\{D_j\}_{j=1}^{d^2}$ are the duals of $\mathcal{S}$ such that $\tr(D^\dagger_i \rho_j) = \delta_{ij}$. The map itself can be represented as a matrix that acts on vectorized density operators as $\Phi=\sum_i \rho_i^\prime \times D_i^*$, where $\times$ indicates an outer product between vectorized quantities.

For any circuit implemented on the IBMQX4, we can perform state tomography on the output and reconstruct via maximum likelihood estimation~\cite{maximum_est, cvxpy, cvxpy_rewriting}. Knowing the input states in $\mathcal{S}$, and hence the duals, gives us sufficient information to compute the map through Eq.~\eqref{eq:tomog}, performing process tomography. Specifically, the full reconstruction signal chain is as follows:
\begin{enumerate}
    \item{Define the $4^N \times 3^N$ unique circuits required given preparation and measurement bases for full process tomography of a desired unitary operation $U$ acting on $N$ qubits.}
    \item{Using the measurement outcome probability distributions, perform maximum likelihood estimation to determine the output state of each circuit.}
    \item{Reconstruct the most likely process given the defined input and reconstructed output states for each circuit, retrieving the map $\Phi_U$. }
\end{enumerate}

As an explicit example of this process for a single qubit operator, one of the $4\times 3 = 12$ unique measurement configurations we would run on the IBMQX4 would be to first create one of the spanning input states; $\KetBra{1}{1}$ in this example. Each qubit in the IBMQX4 is always initialised in the $\KetBra{0}{0}$ state so we first apply a Pauli $X$ to the qubit then apply the operator $U$ to this prepared state. We now wish to know what the output state of $U$ is, given an input of $\KetBra{1}{1}$. Single qubit state tomography requires, at a minimum, measurement of three expectation values -- $\langle X \rangle,\langle Y \rangle,\langle Z \rangle$. The IBMQX4 can only perform projective measurements in the $Z$ basis however and so to determine the others we must first rotate the output state of $U$ before measuring $\langle Z \rangle$. Continuing with our single configuration example, we choose to measure $\langle X \rangle$, requiring a Hadamard $H$ rotation prior to measurement of $Z$. Repeating this many times allows computation of $\langle X \rangle$. Thus we are ultimately measuring $\tr\left(Z HUX\KetBra{0}{0}XU^\dagger H \right)$ as one of 12 measurements to compute the operator $U$, or rather the map $\Phi_U$ when the desired operation $U$ is performed on the IBMQX4.

Since IBM regularly re-calibrates the device, some care does need to be taken when performing the tomography experiments, so as not to mistake fluctuations in system parameters with environmental noise. While different gate combinations were run at different times, the tomography experiments for a particular gate combination $U,V$ are always performed in immediate succession. Additionally, the experiments for each column of Fig.~\ref{fig:errors}b and \ref{fig:errors}c were run sequentially, so that the differences in conditional behaviour for fixed $V$ could be attributed solely to the variation in the preceding choice of gate $U$.

\section{Error contribution analysis}\label{app:B}

For the numerical data presented in Figs.~\ref{fig:errors} and \ref{fig:memory_decay} one can think of the uncertainty in these results stemming from two independent sources. The first being the usual statistical uncertainty for a finite sampling method, while the other is due to state preparation and measurement errors or SPAM errors. First we begin with consideration of the statistical errors.

The output of the IBMQX4 is a probability distribution of the measured states, built from a user defined number of $N$ repeated experiments - the shot number. Assuming independent experiments, the standard deviation of the probabilities for each measured outcome goes as $\delta = \tfrac{1}{\sqrt{N}}$. Since we assume no measurement error bias, this uncertainty defines a sphere (truncated by the space of valid quantum states) in $\mathcal{B}(\mathcal{H})$ with the true state being the center and the radius being the standard deviation $\delta$. Our analysis methods from this point involve taking this outcome distribution and estimating the most probable quantum state, then using this state tomography in process tomography which is then used for comparing the temporal dependence of different quantum processes \'{a} la Figs.~\ref{fig:errors}, \ref{fig:conditionals}, and \ref{fig:memory_decay} . By using a maximum likelihood estimator it is not immediately clear how to propagate the statistical uncertainty. We can however simulate this process numerically and gain, at the very least, a lower bound on the uncertainty. By sampling from the Gaussian distributed states, with mean centered on the true state, we can propagate a large number of measured states through our analysis process and compute a distribution of values for the entries in Figs.~\ref{fig:errors}, \ref{fig:conditionals}, and \ref{fig:memory_decay}. The standard deviation of the  distribution of values found via this large scale sampling of quantum states then indicates the standard deviation in our output values due to a finite experiment, giving a bound on the standard deviation in the values of Fig.~\ref{fig:errors}, \ref{fig:conditionals} and \ref{fig:memory_decay}.

Suppose for the moment the preparation and measurement errors are in fact zero and there is only a difference in the map we think we are applying (the ideal unitary) and the actual. The true process run in the noisy IBMQX4 then might be expressed as

\begin{gather}\label{eq:tchannel}
\Phi_U = \Phi_U^{\rm id} + \alpha_U,
\end{gather}
with $ \Phi_U^{\rm id} $ being the ideal process we wish to apply and $\alpha_U$ being an error channel that introduces unwanted perturbations in the output states of $\Phi_U$. It is the behaviour of $\alpha_U$ and its dependence on the history of the system--environment that we have been concerned with in this paper. Our hope and assumption up until now is that the channel found through tomography
\begin{gather}\label{eq:beta}
\Phi_U^{\prime} = \Phi_U^{\rm id} + \beta_U,
\end{gather}
is exactly \eqref{eq:tchannel}. For perfect measurement/preparation channels the tomography error channel $\beta_U$ is equivalent to $\alpha_U$ and the reconstructed channel and the actual are identical. Since in any real experiment this will not be true, we wish to know the error on the error itself
\begin{gather}\label{eq:error}
e(U) =   \Phi_U - \Phi_U^{\prime} =  \alpha_U - \beta_U.
\end{gather}
We shall argue that even with the measurement errors accounted for, our results remain significant.

Process tomography is a three step process; state preparation, evolution and measurement. The preparation stage involves constructing an input set of states $\rho_i$ that span the space of bounded linear operators $\mathcal{L}(\mathcal{H}_d)$ that $\Phi_U$ acts on. If one knows how a map acts on a complete spanning set the full process for an arbitrary input state may be derived. 
The IBMQX4 is always well initialized as $\rho = \KetBra{0}{0}^{\otimes 5}$ thus the preparation stage is a set of operations that map this initial state to an element in the spanning set. Since these maps will invariably have some error associated with them, the actual prepared states are $\{\rho_i\} = \{\Phi_i[\KetBra{0}{0}] + \gamma_i [\KetBra{0}{0}] \}$. If the actual input states to the process $\Phi_U$ has some error, then so too do the duals $\{D_i^\prime \} = \{D_i + \delta_i\}$ as they are constructed based off an assumed spanning set with $\{D_i \}$ being the actual duals to $\{ \rho_i \}$.

When the input state drawn from $\{\rho_i \} $ is passed through the process $\Phi_U$ the channel can be thought of as acting on the desired input state and the error introduced by the preparation stage, with the output states of the channel becoming
\begin{gather}
    \rho_i^\prime  = \Phi_U \circ \Phi_i [\KetBra{0}{0}] +  \Phi_U \circ \gamma_i [\KetBra{0}{0}]  .
\end{gather}
The set $\{\rho_i^\prime \}$ is determined through state tomography, but, since we are restricted to purely $Z$ projective measurements in the IBM system, measurements in the remaining elements of the Pauli basis (or any other measurement basis for that matter) first require the application of a rotation operator on the output states of $\{\rho_i^\prime \}$ before projective measurement. This rotation in the measurement stage introduces another source of errors beyond simple statistical uncertainty. This is to say that we measure not $\{\rho_i^\prime \}$ but another output state that is the output of $\Phi_U^\prime$ plus the error $\eta_i$ introduced by imperfect rotation operators
\begin{gather}
    \rho_i^{\prime \prime} = \rho_i^\prime + \eta_i[\rho_i^\prime].
\end{gather}

If we then compute the channel defined by this new set of output states and the duals of the input states using Eq.~\eqref{eq:tomog} we find an expression for the reconstructed channel in terms of the actual channel and the error terms; 
\begin{equation}
\begin{split}
    \Phi_U^{\prime } & = \sum_i \rho_i^{\prime \prime} \times D_i^{\prime *},\\
    & =\sum_i \left(\rho_i^\prime + \eta_i[\rho_i^\prime] \right) \times \left( D_i^* + \delta_i^* \right)
\end{split}
\end{equation}
If we then substitute into Eq.~\eqref{eq:error} terms of compositions of the channels and simplify, we arrive at an expression for the error $e(U)$
\begin{align*}
    e(U) = & \sum_i \left(\Phi_U\circ \gamma_i \left[ \KetBra{0}{0} \right] + \eta_i \circ \Phi_U \circ \Phi_i \left[ \KetBra{0}{0} \right]  \right) \times D_i^* \\
    & + \Phi_U \circ \Phi_i \left[ \KetBra{0}{0} \right]\times \delta_i^\prime + \mathcal{O}(\eta_i \circ \gamma_i, \eta_i\circ \delta, \gamma_i \circ \delta_i).
\end{align*}

Considering only terms up to first order in the errors, we see that the SPAM errors at each stage of the tomography each contribute an independent error term. Ignoring the second order errors, the first order preparation/measurement errors are due solely to the imperfection of the preparation/measurement channels with the finite sampling contributing to the latter as well. Previously we argued that the finite sampling is a small contribution and by IBM's own gate analysis (using randomised bench-marking), and available as part of the IBMQX4 interface, we know that the gates used for preparation are projective measurement are small as well. This lets us conclude that each term in the above equation is small at first order and thus so to is $e(U)$, indicating that the results presented in the main text are due to non-Markovian behaviour rather than a simple infidelity.

\begin{acknowledgements}
We thank Nathan Langford and Chris Wood for insightful conversations. KM is supported through Australian Research Council Future Fellowship FT160100073.
\end{acknowledgements}

\bibliography{IBMreference}
\end{document}